\documentstyle[12pt,epsfig]{article}
\renewcommand{\bar}[1]{\overline{#1}}

\newcommand{\optbar}[1]{\shortstack{{\tiny (\rule[.4ex]{1em}{.1mm})} 
  \\ [-.7ex] $#1$}}

\global\arraycolsep=2pt %reduces the separation in eqnarrays

\newcommand{\AmS}{{\protect\the\textfont2
 A\kern-.1667em\lower.5ex\hbox{M}\kern-.125emS}}

\newread\epsffilein % file to \read
\newif\ifepsffileok % continue looking for the bounding box?
\newif\ifepsfbbfound % success?
\newif\ifepsfverbose % report what you're making?
\newdimen\epsfxsize % horizontal size after scaling
\newdimen\epsfysize % vertical size after scaling
\newdimen\epsftsize % horizontal size before scaling
\newdimen\epsfrsize % vertical size before scaling
\newdimen\epsftmp % register for arithmetic manipulation
\newdimen\pspoints % conversion factor
\def\epsfbox#1{\global\def\epsfllx{72}\global\def\epsflly{72}%
 \global\def\epsfurx{540}\global\def\epsfury{720}%
 \def\lbracket{[}\def\testit{#1}\ifx\testit\lbracket
 \let\next=\epsfgetlitbb\else\let\next=\epsfnormal\fi\next{#1}}%
\def\epsfgetlitbb#1#2 #3 #4 #5]#6{\epsfgrab #2 #3 #4 #5 .\\%
 \epsfsetgraph{#6}}%
\def\epsfnormal#1{\epsfgetbb{#1}\epsfsetgraph{#1}}%
\def\epsfgetbb#1{%
\openin\epsffilein=#1
\ifeof\epsffilein\errmessage{I couldn't open #1, will ignore it}\else
 {\epsffileoktrue \chardef\other=12
 \def\do##1{\catcode`##1=\other}\dospecials \catcode`\ =10
 \loop
 \read\epsffilein to \epsffileline
 \ifeof\epsffilein\epsffileokfalse\else
 \expandafter\epsfaux\epsffileline:. \\%
 \fi
 \ifepsffileok\repeat
 \ifepsfbbfound\else
 \ifepsfverbose\message{No bounding box comment in #1; using defaults}\fi\fi
 }\closein\epsffilein\fi}%
\def\epsfclipstring{}% do we clip or not? If so,
\def\epsfsetgraph#1{%
 \epsfrsize=\epsfury\pspoints
 \advance\epsfrsize by-\epsflly\pspoints
 \epsftsize=\epsfurx\pspoints
 \advance\epsftsize by-\epsfllx\pspoints
 \epsfxsize\epsfsize\epsftsize\epsfrsize
 \ifnum\epsfxsize=0 \ifnum\epsfysize=0
 \epsfxsize=\epsftsize \epsfysize=\epsfrsize
 \epsfrsize=0pt
 \else\epsftmp=\epsftsize \divide\epsftmp\epsfrsize
 \epsfxsize=\epsfysize \multiply\epsfxsize\epsftmp
 \multiply\epsftmp\epsfrsize \advance\epsftsize-\epsftmp
 \epsftmp=\epsfysize
 \loop \advance\epsftsize\epsftsize \divide\epsftmp 2
 \ifnum\epsftmp>0
 \ifnum\epsftsize<\epsfrsize\else
 \advance\epsftsize-\epsfrsize \advance\epsfxsize\epsftmp \fi
 \repeat
 \epsfrsize=0pt
 \fi
 \else \ifnum\epsfysize=0
 \epsftmp=\epsfrsize \divide\epsftmp\epsftsize
 \epsfysize=\epsfxsize \multiply\epsfysize\epsftmp
 \multiply\epsftmp\epsftsize \advance\epsfrsize-\epsftmp
 \epsftmp=\epsfxsize
 \loop \advance\epsfrsize\epsfrsize \divide\epsftmp 2
 \ifnum\epsftmp>0
 \ifnum\epsfrsize<\epsftsize\else
 \advance\epsfrsize-\epsftsize \advance\epsfysize\epsftmp \fi
 \repeat
 \epsfrsize=0pt
 \else
 \epsfrsize=\epsfysize
 \fi
 \fi
 \ifepsfverbose\message{#1: width=\the\epsfxsize, height=\the\epsfysize}\fi
 \epsftmp=10\epsfxsize \divide\epsftmp\pspoints
 \vbox to\epsfysize{\vfil\hbox to\epsfxsize{%
 \ifnum\epsfrsize=0\relax
 \includegraphics{#1}%
 \else
 \epsfrsize=10\epsfysize \divide\epsfrsize\pspoints
 \includegraphics{#1}%
 \fi
 \hfil}}%
\global\epsfxsize=0pt\global\epsfysize=0pt}%
 {\catcode`\%=12 \global\let\epsfpercent=%\global\def\epsfbblit{%BoundingBox}}%
\long\def\epsfaux#1#2:#3\\{\ifx#1\epsfpercent
 \def\testit{#2}\ifx\testit\epsfbblit
 \epsfgrab #3 . . . \\%
 \epsffileokfalse
 \global\epsfbbfoundtrue
 \fi\else\ifx#1\par\else\epsffileokfalse\fi\fi}%
\def\epsfempty{}%
\def\epsfgrab #1 #2 #3 #4 #5\\{%
\global\def\epsfllx{#1}\ifx\epsfllx\epsfempty
 \epsfgrab #2 #3 #4 #5 .\\\else
 \global\def\epsflly{#2}%
 \global\def\epsfurx{#3}\global\def\epsfury{#4}\fi}%
\def\epsfsize#1#2{\epsfxsize}
\let\epsffile=\epsfbox
\def\sss{\scriptscriptstyle}
\def\barp{{\raise.35ex\hbox{${\sss (}$}}---{\raise.35ex\hbox{${\sss )}$}}}
\def\bdbarp{\hbox{$B_d$\kern-1.4em\raise1.4ex\hbox{\barp}}}
\def\bsbarp{\hbox{$B_s$\kern-1.4em\raise1.4ex\hbox{\barp}}}
\def\dbarp{\hbox{$D$\kern-1.1em\raise1.4ex\hbox{\barp}}}
\begin{document}
%\begin{flushright}
%DESY 98-xxx\\
%hep-ph/9805yyy \\
%May 1998\\
%\end{flushright}
\begin{center}
{\Large \bf
\centerline{Quark Mixing and CP Violation}}
\vspace*{1.0cm}
%\vskip1cm
{\large Ahmed Ali}
\vskip0.2cm
Deutsches Elektronen-Synchrotron DESY \\
Notkestra\ss e 85, D-22603 Hamburg, FRG\\
\vskip0.5cm
{\large Boris Kayser}
\vskip0.2cm
Division of Physics, National Science Foundation, \\
4201 Wilson Boulevard, Arlington, VA 22230, USA\\
\vskip1.0cm
\begin{abstract} 
 We review the historical developments leading to the
Cabibbo-Kobayashi-Maskawa (CKM) quark mixing matrix. Present
status of the CKM matrix from direct measurements is summarized,
giving also the present profile of the unitarity triangle. CP Violation
in the $K^0$ - $\overline{K^0}$ complex and in $B$-meson decays are
discussed in the context of the CKM matrix. 
 \end{abstract}
\vspace*{2.0cm}
   Invited article; to be published in `The Particle Century', Institute of
Physics Publishing Inc., Bristol and Philadelphia, (1998);\\
Editor: Gordon Fraser.
\end{center}

\thispagestyle{empty}
\newpage
\setcounter{page}{1}

\section{Quark Flavour Mixing}
%\section{The Cabibbo-Kobayashi-Maskawa Matrix}

Elementary particles carry many additive attributes (quantum
numbers) which are conserved in the strong and electromagnetic
interactions.
These quantum numbers are called flavours and are  
used to characterize hadrons (strongly interacting particles).
If electromagnetism and strong nuclear forces were the only interactions
in nature, there would have been no flavour changing reactions seen
in laboratory experiments. However, it has been known since the early days of
weak interactions that the neutron is unstable and it decays into a proton
by emitting an electron and its associated antineutrino $n \to p e^- 
\overline{\nu_e}$ with a mean life of about 15 minutes.
On the other hand, to date not a single proton decay has been observed.
Laboratory experiments have put the proton lifetime in excess of $10^{32}$ years,
which is some 22 orders of magnitude larger than the age of our universe!

In  quark language, the two lightest quarks, called up (or $u$),
having the fractional electric charge $+2/3$,
and down (or $d$), having the fractional electric charge $-1/3$
are at the base of the neutron beta (electron-emitting) decay and the stability
of the proton. One can imagine that the u and d quarks form a doublet and the
charged weak interaction causes a transition from the heavier $d$ to the lighter
$u$ component. Then, a neutron, which consists of two $d$ and one $u$ quarks
($n=ddu)$ turns into a proton $(p=uud)$, which consists of two $u$ and one $d$
quarks with the charged
weak interaction causing the transition $d \to u e^- \overline{\nu_e}$.
The lightest of the quarks, the $u$ quark, is then stable, as
ordinary charged weak interactions do not allow the transition of a quark
into a lepton. The consequence of this is that a 
proton, being the lightest known baryon (a hadron with spin $1/2$),
remains stable. This example shows that    
charged weak interactions allow transitions between hadrons (or quarks)  
with different flavour quantum numbers. Here the quark flavours are up
and down. 

Back in 1933, Enrico Fermi wrote
an effective (i.e, low energy) theory of charged weak interactions by 
introducing an effective
coupling constant $G_F$, the Fermi coupling constant. In the example given above,
the mean lifetime of the neutron determines the strength of $G_F$ and present 
day experiments have measured it very precisely: 
$G_F= 1.166392(2) \times 10^{-5}$ GeV$^{-2}$ in units used by
particle physicists in which the reduced Planck constant $h/2 \pi$ and
the velocity of light are both set to unity.
There are other known reactions in which charged weak interactions are at
work. An example is the decay of a muon into an electron, a neutrino and an
antineutrino, $\mu^- \to e^- \overline{\nu_e} \nu_{\mu}$. The decay rate,
hence the lifetime of the muon, is also determined by the Fermi
coupling constant. For a long time, until 
experimental precision improved, it was generally accepted that
the Fermi coupling constants in neutron beta decay, 
$G_n$, and in the muon decay, $G_\mu$, were one 
and the same. However, as the experimental precision
improved and theoretical calculations became more sophisticated,
by including quantum corrections as well as nucleus-dependent 
effects in nuclear beta decays, from where most of the information on
neutron beta decay comes, it was established that indeed $G_n \neq G_\mu$,
though the difference is small. Today,
thanks to very precise experiments, this difference is known
very precisely: $G_n/G_\mu = 0.9740 \pm 0.001$.
So, it seemed that experiments on nuclear beta decay and muon decay required
not {\it one} but {\it two} different Fermi coupling constants.

As the particle zoo enlarged, in particular with the discoveries of hadrons which
carry a new quantum number called strangeness, it became clear that the decay
rates of these newly discovered weakly decaying particles were different. A
successful description of the decay widths (a measure of transition rate) of
kaons and hyperons (nucleon-like particles with a non-zero strangeness quantum
number) required introduction of effective coupling constants
which were very different than either $G_n$ or $G_\mu$. A good example
is the decay of a charged kaon, $K^-$, which was found to decay into a
neutral pion $\pi^0$, an electron and an electron-anti neutrino, $K^- \to 
\pi^0 e^- \bar{\nu}_e$.
In this case, the effective Fermi coupling constant was found to have an
empirical value of $G_K/G_\mu \simeq 0.22$. In quark language, this
transition is induced by the mutation $s \to u e^- \bar{\nu}_e$, as the charged
kaon has the quark content $K^- =s\bar{u}$ and a neutral pion is built
up from the linear combination of the up and down quarks and their antiquarks $\pi^0 
=1/\sqrt{2}(u\bar{u} - d\bar{d})$,
reflecting its isospin properties. So, experiments seemed to have implied
the existence of at least {\it three} different Fermi coupling constants,
$G_n$, $G_\mu$ and $G_K$. 
The question in the theory of weak interactions being asked in the early 
sixties was:  Should one give up the concept of a universal charged weak
interaction, as opposed to electromagnetism and the strong nuclear force? 

The answer came in the hypothesis of flavour mixing, implying that
the quantum eigenstates of the charged weak interactions are rotated in
quark flavour space with respect to the mass eigenstates. In other words, the 
states which have simple charged weak interactions are not the states of 
definite mass, but linear combinations of them. The concept of `rotated' 
charged weak currents
(involving the $W$ bosons) in the flavour space
was introduced by Nicola Cabibbo in 1963, following an earlier suggestion
by Murray Gell-Mann and Maurice Levy. 
It solved the two outstanding problems in weak interactions, explaining the
strongly suppressed weak decays of the kaons and hyperons
compared to the weak decays of the non-strange light hadrons
(containing the $u$ and $d$ quarks), and the difference
in the strength of the nuclear $\beta$-decays compared to
$\mu$-decay. Calling the Fermi coupling constant in $\mu$-decay
$G_F$, the coupling constants in neutron
$\beta$-decay and the strange hadron decays in the   
Cabibbo theory are given by $G_F\cos \theta_C$ and
$G_F\sin\theta_C$, respectively.  Here, the Cabibbo angle $\theta_C$
is the angle between the weak eigenstate and the mass
eigenstate of the quarks.  A value $\theta_C\simeq$ 13 degrees describes all 
data involving weak decays of light hadrons with the same Fermi coupling
constant, preserving the universality of weak interactions.   

The rates for numerous weak transitions
involving light hadrons or leptons are adequately accounted for
in the Cabibbo theory in terms of two quantities, $G_F$ and $\theta_C$.
This was a great triumph.
However, Cabibbo rotation with three light quarks turned out to cause 
havoc for  so-called 
flavour-changing-neutral-current (FCNC) processes, which in this theory 
were not in line with their effective measured strengths.

What are these FCNC processes? One example from the Cabibbo epoch characterizes
the processes in question. Consider particle-antiparticle mixing involving the
neutral kaon ($K^0$ - $\overline{K^0}$) complex, in which, through a virtual
transition, a $K^0(=\bar{s}d)$ meson turns into its charge conjugate antiparticle
$\overline{K^0}(=s\bar{d})$. Now, the $K^0$ meson has the strangeness quantum
number $S$ equal to $+1$. The strangeness quantum number of its conjugate
antiparticle $\overline{K^0}$ is then $S=-1$. So, in the virtual
$K^0-\overline{K^0}$ transition, the electric charge $Q$ does not change, i.e.,
in this process $\Delta Q=0$, but $S$ has changed by two units, i.e., $\Delta
S=2$. Such transitions, and we shall see several counterparts in heavy meson
systems, are FCNC processes. 

Since the quantum number $S$ is conserved in
strong and electromagnetic interactions, the $K^0-\overline{K^0}$ states can 
not be mixed by these forces.
The charged weak force is the only known force which changes flavours,
so it must be at work in inducing the $K^0-\bar{K^0}$ mixing. Now, it is 
known that mixing of two degenerate levels must result in level splitting,
introducing mass differences between the mass eigenstates, named 
$K_S$ and $K_L$, being 
the short-lived and longer-lived of the two mesons.
The mass difference $\Delta M_K \equiv M(K_L) - M(K_S)$ in the Cabibbo 
theory turned out to be several orders of magnitude larger than the observed 
mass difference, whose present day value is $\Delta M_K \simeq 3.49 \times 
10^{-6}$ eV. (As a fraction of $M_K$, the average of the $K_L$ and $K_S$ masses, 
this mass difference is only $7.0 \times 10^{-15}$!)

\begin{figure}[htb]
\vskip -0.4truein
\centerline{\epsfysize=7in
{\epsffile{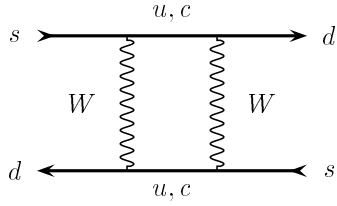}}}
\vskip -5.0truein
\caption{The box diagram contributing to the mass difference $\Delta M_K$
in the GIM theory. In the six quark theory, also the top quark contributes
whose contribution is small and hence not shown.}
\label{boxdiagram}
\end{figure}

This great disparity in the effective strength 
of the $K^0-\bar{K^0}$ transition in the Cabibbo theory and experiment
remained for a long time a stumbling block in developing a consistent 
theory of neutral weak currents involving hadrons. For example, it was
not at all obvious if the same weak force which causes
the decays of the muon, neutron, and the charged kaon discussed above was at 
work in $K^0-\overline{K^0}$ transition, or whether a new effective
force had to be invented to explain $\Delta M_K$. During   
this epoch came the seminal papers by
Steven Weinberg and Abdus Salam in 1967/68, proposing renormalizable 
weak interaction models for leptons unifying weak and electromagnetic 
interactions (see Chapter Rubbia), in which the outstanding problem of
FCNC hadronic weak interactions was pushed to one side.
It took several years after the advent of this electroweak theory before the 
FCNC problem was solved elegantly through the 
`GIM'  mechanism, invented by Sheldon Glashow, John Iliopoulos and Luciano
Maiani in 1970, using the hypothesis of the fourth (charm or simply c) quark.
According to the GIM proposal the charge-changing ($W$-emitting) weak
current involving quarks has the form
\begin{equation}
J = \bar ud^\prime +\bar cs^\prime \ ,
\label{eqA1}
\end{equation}
where $d^\prime$ and $s^\prime$ are rotated (orthogonal) combinations of the
$d$ and $s$ quarks which can be described in terms of the Cabibbo angle
$\theta_C$ as:
\begin{eqnarray}
d^\prime &=& d\cos\theta_C+s\sin\theta_C \nonumber\\
s^\prime &=& -d\sin\theta_C+ s\cos\theta_C \ .
\label{eqA2}
\end{eqnarray}
The GIM construction of the charge-changing weak current,   
involving four quark flavours $(u,d,s,c)$, removed the leading
contribution to the $K_L$-$K_S$ mass difference. Quantum 
(loop) effects, such as the ones shown in the box diagram of
Fig.~\ref{boxdiagram}, with the contribution of the $u$ and $c$ quarks
in the intermediate states, give nonzero contributions
to the $K^0-\overline{K^0}$  mass difference. The result of the box
diagrams can be written as (here $m_\mu$ is the mass of the muon)
\begin{equation}
  \Delta M_K \simeq \frac{4 (m_c^2 - m_u^2) \cos^2 \theta_C}{3 \pi m_\mu^2}
  \Gamma(K^+ \to \mu^+ \nu_\mu)~.
\end{equation}

%
% This is Figure 1
%
 
With the other quantities known, $\Delta M_K$ could be predicted in terms
of the mass difference of the charm and up quark. This led 
Benjamin Lee and Mary Gaillard in 1972 to estimate
a mass of $1$ - $2$ GeV for the charm quark in the Cabibbo-GIM four-quark
theory. The GIM proposal remained a curiosity until the
charm quark was discovered in 1974 by the experimental groups led by 
Samuel Ting, at the Brookhaven National Laboratory, and Burt Richter,
at the Stanford Linear Accelerator SLAC,  
through the $c\bar c$ bound state $J/\psi$ (see Chapter Schwitters), with the
charm quark mass compatible with theoretical estimates.
Subsequent discoveries of the charmed hadrons $D^0 (=c\bar{u})$,
$D^+(=c\bar{d})$, $D_s(=c\bar{s})$ at SLAC, DESY and elsewhere and
their weak decays have confirmed the Cabibbo-GIM current, with again all the
decays governed essentially by the parameters $G_F$ and $\theta_C$,
thus restoring the universality of charged weak interactions.

We now know that there are not four but six quarks. The charged weak currents
then would involve linear combinations of these quarks. We leave 
the discussion of flavour mixing in 
the six quark theory to a subsequent section and discuss first another development
in kaon decays which had a profound effect on the theoretical
developments, namely CP violation.  
\section{CP Violation in the $K^0$ - $\overline{K^0}$ Complex}

For every elementary particle, there is a corresponding
antiparticle.  However, a particle and its antiparticle do not
always behave in the same way.  For example, in the process
$\pi^+\rightarrow \mu^++\nu$, in which a positively charged pion
decays into a positively charged muon and a neutrino, the muon
emerges with its spin vector {\it antiparallel} to  its momentum.
By contrast, in the process $\pi^-\rightarrow \mu^-+ \bar\nu$, in
which every particle in the previous decay has been replaced by its
antiparticle, the muon emerges with its spin {\it parallel} to its
momentum.  This difference between the two processes shows that the
world is not invariant under charge conjugation $C$, which replaces
every particle by its antiparticle.  However, one may wonder whether
the world is nevertheless unaltered by matter-antimatter interchange
in the sense that it is invariant under charge conjugation {\it
combined with a parity (space) reflection P}.  Under $P$, particle spins
do not change, but their momenta are reversed.  Thus, the CP-mirror image
of the process $\pi^+\rightarrow \mu^+$ (spin antiparallel to
momentum) $+~ \nu$ is $\pi^-\rightarrow \mu^-$ (spin parallel to
momentum) $+~ \bar \nu$.  Experimentally, the rates for these two
processes are equal.  Thus, CP invariance holds for $\pi
\rightarrow \mu\nu$. 

 However, it has been discovered that CP does
not hold everywhere. There are, as already noted, two neutral $K$ mesons,
the short-lived $K_S$ (decaying mainly into two pions) and the longer-lived 
$K_L$ (decaying mainly into $\pi e \nu$,$ \pi\mu\nu$, or three pions).  If CP 
invariance held, $K_S$ and $K_L$ would each be its own CP-mirror image.
Thus, the CP-mirror image of the decay $K_L\rightarrow \pi^-e^+\nu$
would be the decay $K_L\rightarrow \pi^+e^-\bar\nu$, with all the
momenta in the first decay reversed.  If we ask about the rates for
these two decays integrated over all possible directions of the
outgoing particles, the momentum reversals become irrelevant, and CP
invariance would require that the two rates be equal.  However,
these rates differ by 0.3\%.  Thus, the world is noninvariant, not
only under $C$, but under CP as well.  Noninvariance
under the symmetry operation CP is accompanied by nonconservation
of the associated CP quantum  number, and  the first observation
that either of these phenomena occurred was the discovery in 1964 that the CP
quantum number is not conserved in the decays of neutral $K$ mesons
to pairs of pions.  The CP quantum number of a system, referred to
as its CP parity, can be either $+1$ or $-1$.  
If CP were conserved,  $K_S$ and $K_L$ would be CP
eigenstates with opposite CP parity.  The
CP parity of the pion pair $\pi^+\pi^-$ (the dominant decay of the $K_s$) is 
even. However, in 1964 it was discovered by J. Christenson, James
Cronin, Val Fitch, and Ren\'e Turlay that the process
$K_L\rightarrow \pi^+\pi^-$ also occurs. That is, both $K_S$ and $K_L$, one
 of which would have CP $= -1$ if CP were conserved, decay to $\pi^+\pi^-$,
 which has CP $= +1$. Thus CP is not conserved
in neutral $K$ meson decays, although the observed
nonconservation is small: the ratio of the CP-violating amplitude to the
 CP-conserving one,
 $|{\rm Amp}(K_L\rightarrow \pi^+\pi^-)/{\rm Amp}(K_S\rightarrow
\pi^+\pi^-)|$, is only $2.3\times 10^{-3}$.  However, much larger
effects may be revealed in the future, as we shall see.

Now, CP violation has so far been seen only in the decays of neutral $K$
mesons.  Thus, this violation could perhaps be a feature of
$K^0-\bar{K^0}$ mixing, rather than of particle decay amplitudes.
Then there would be no CP violation in the decays of {\it charged}
$K$ mesons.  (The charged $K$ mesons, $K^+$ and $K^-$, certainly do
not mix, because the conversion of one of them into the other would
violate charge conservation.)  Several very challenging
experimental efforts are being made to see whether decay amplitudes
do violate CP.  So far, the results are inconclusive.  One
experiment finds that the quantity
``$\epsilon^\prime_K/\epsilon_K$,'' whose nonvanishing would signal
that the decay amplitudes violate CP, is $(23.0\pm6.2)\times
10^{-4}$, but another finds that it is $(7.4\pm 5.9)\times 10^{-4}$,
consistent with zero. The experiments continue.

Regardless of the value of $\epsilon^\prime_K/\epsilon_K$, the fact
that nature violates CP invariance and CP conservation has been
established.  What is the origin of this CP violation?  In
addressing this question, we note that, as remarked earlier,
CP-violating effects have thus far been observed only in the decays
of neutral $K$ mesons.  These decays are known to be due to the weak
interaction.  Thus, it is natural to ask whether CP violation is
also due to the weak interaction, rather than to some so-far
unknown, mysterious force.

Among the discrete symmetries
C, CP, T, and CPT, the CPT symmetry is considered to be exact as it follows
from fundamental principles underlying all field
theories, namely positivity of the norm and locality (a particle is 
represented by a local field). Lately, the invariance of natural
laws under CPT has been put in question in the context of the superstring
theories of particle physics, in which the particles are described by
extended objects in space-time, such as a string, lifting the assumption
of locality (point-like nature) ascribed to the particles in field
theories (see Chapter Ross). However, even if CPT invariance should prove
 to be broken in superstring theory, the effects on the flavour physics
 being discussed in this Chapter would very likely be negligible. Therefore,
 in what follows, we shall assume that CPT holds exactly. The CPT-invariance
 principle has a number of implications, such
as the equality of the masses and of the lifetimes of a particle and its
antiparticle. The best limit on CPT violation stems from the upper limit of
the ratio of mass difference to the mass,
 $m(K^0) - m(\overline{K^0})/m(K^0) \leq 10^{-18}$. 

If CP violation is indeed produced by the known weak interaction
described by the Standard Model, then it is caused by complex phase
factors in the quark mixing matrix.  How  these complex phases
produce physical CP-violating effects will be explained shortly.

\section{The Cabibbo-Kobayashi-Maskawa Matrix}

We now know  that there are six quarks in nature.
The fifth and the sixth quarks are called the beauty (or simply $b$)
and top (or simply $t$) quarks. The $b$ quark was discovered in the form    
of its bound state $\Upsilon=(b\bar{b})$ and excited states
$\Upsilon^\prime,... $ by the group of Leon Lederman working at Fermilab 
in 1977. The discovery of the top quark had to wait until
1994 when two large experimental groups (D0 and CDF) working again at
Fermilab finally discovered the top quark  in the process
$p\bar{p} \to t \bar{t}X$ and the subsequent decays of the top quarks
$t \to bW^+$ (see Chapter Shochet).
However, indirect evidence of a top quark with a rather large mass,
$m_t = O(100)$
GeV, was found earlier from $B^0$ - $\overline{B^0}$ mixing
by the UA1 experiment at CERN's proton-antiproton collider (see Chapter 
Rubbia), the ARGUS experiment at DESY's DORIS electron-positron collider, and 
the CLEO experiment at the Cornell electron-positron collider CESR. The
 expectation that $m_t$ is large was further strengthened
by electroweak precision measurements at CERN's LEP. No top meson has so far
been constructed from its decay product, but there exists an impressive amount
of data on the properties of the beauty hadrons from experiments at
DORIS (DESY), CESR (Cornell), LEP (CERN), SLAC (Stanford) and Fermilab. 
  
Given that there are six quarks, arranged in terms of three ``weak isospin''
doublets $(u,d;~s,c;~t,b)$,
the obvious questions are: How are the weak interaction 
eigenstates involving six quarks related to the quark mass eigenstates?
And what does this rotation imply for the decays of the light (containing  
only $u,d,s$ quarks) 
and heavy hadrons (containing $c,b,t$ quarks)?
% Can the observed CP violation
%in the $K^0$ - $\overline{K^0}$ system be quantitatively understood?

The answers follow if the two-dimensional
rotation of Eq.~(\ref{eqA2}) is now replaced by a three-dimensional
one, where the $W$-emitting weak current takes the form
\begin{equation}
J=\bar ud^\prime +\bar cs^\prime +\bar tb^\prime
 =(\bar u, \bar c,\bar t)\, V
\left(\begin{array}{c} d\\s\\b\end{array}\right) \ .
\label{eqA4}
\end{equation}
Here $V$ is a $3\times 3$ matrix in the quark flavour space and can be
symbolically written as:
\begin{equation}
V = \left(\begin{array}{ccc}
V_{ud} & V_{us} & V_{ub} \\
V_{cd} & V_{cs} & V_{cb} \\
V_{td} & V_{ts} & V_{tb} \end{array}\right) \ .
\label{eqB1}
\end{equation}
Thus, in this case the weak (interaction) eigenstate $d^\prime$ is:
$d^\prime \equiv V_{ud}d + V_{us}s+V_{ub}b$. (Likewise, the other rotated
states $s^\prime$ and $b^\prime$ follow from eq.~(\ref{eqB1}).)
Every weak process involving the W boson is proportional to some product of
 the elements of V. Now comes a crucial observation: if some of the elements 
of the $3\times3$ matrix $V$ are not real but complex (so that $V$ is not,
strictly speaking, a rotation matrix, but a ``unitary'' one), then the  
hadronic weak interactions can violate CP.  
However, if there were only 
four quarks, so that we did not have a $3\times3$ quark mixing matrix but
only the two-dimensional rotation of Eq. (\ref{eqA2}) of the Cabibbo-GIM
theory, then  making the coefficients in that rotation complex would not 
lead to any physical effects, as in this case these complex phases in
the $(2\times 2)$ rotation matrix can be eliminated by a 
redefinition of the quark fields. Not so, if there are six or more quarks.

These facts were first pointed out by M.~Kobayashi and T.~Maskawa in 1972, long
before the $c$, $b$, and $t$ quarks were actually discovered.
In fact, the GIM mechanism, put forward to describe FCNC transitions
in the $K^0-\bar{K^0}$ complex, was immediately followed by the KM 
hypothesis to accommodate CP violation in the same $K^0$-$\overline{K^0}$
system and which predicted the existence of all the three heavy quarks.

Now that all these quarks have been discovered, the six
quark theory of Kobayashi and Maskawa stands on firm experimental
ground - as far as its quark content is concerned. The crucial question now 
is whether the complex phases in the matrix $V$ are the only source of CP
 violation in flavour physics. These phases predict CP-violating phenomena
 in the decay amplitudes (also called {\it direct CP violation}) of many 
hadrons. They also predict {\it indirect
CP violation}, which resides in the mass matrix of a neutral meson complex,
$M^0$ - $\overline{M^0}$, in particular $K^0$ - $\overline{K^0}$, and can
manifest itself only when such mixings are involved. It is widely appreciated
 that 
$B$ physics has the potential of providing crucial tests of the KM paradigm.   
The $3\times 3$ flavour mixing matrix, which is now aptly called  the 
Cabibbo-Kobayashi-Maskawa (CKM) matrix, plays a central role in 
quantifying CP-violating asymmetries.
\subsection{Present status of the CKM Matrix}

The magnitudes of all nine elements of the CKM matrix have now
been measured in weak decays of the relevant quarks, and in some cases
in deep inelastic neutrino nucleon scattering. The precision on these
matrix elements
varies for each entry, reflecting both the present experimental 
limitations but
often also theoretical uncertainties associated with the imprecise knowledge
of the hadronic quantities required for the analysis of data.
 In most cases, the decaying particle is a hadron and not
a quark and one has to develop a prescription for transcribing the 
simple quark language to that involving hadrons. Here, the theory of strong
interactions, Quantum Chromodynamics (QCD), comes to the rescue. Powerful
calculational techniques of QCD, in particular renormalization group 
methods, lattice-QCD, 
QCD sum rules and effective heavy quark theory, have been used to
estimate, and in some cases even determine, the hadronic quantities of interest, 
thereby reducing theoretical errors on the CKM matrix elements.

This theoretical development is very impressive though the QCD technology has
not quite reached its goal of achieving an accuracy of a few percent in the
determination of the hadronic matrix elements. Nevertheless, it has
been crucial in quantifying the CKM matrix elements. Fascinating as these
calculational aspects are, their 
discussion here would take us far from our mainstream interest and we 
refer to the suggested literature for further reading.
 
 Present knowledge of $V$ comes from a variety of different
sources and the present status can be summarized as follows:
\begin{equation}
  |V| = \left(\begin{array}{ccc}
  0.9730 - 0.9750 & ~~~0.2173 - 0.2219 & ~~~0.0023 - 0.0040 \\
  0.208 - 0.24& 1.20 - 0.88 & ~0.038 - 0.041 \\
  0.0065 - 0.0102 & 0.026 - 0.040 & 1.14 - 0.84
 \end{array}\right) \ .
\label{eqvckmnum}
\end{equation}

The following comments about the entries are in order:

(1) $|V_{ud}|$: This is based on comparing nuclear beta decays
$(A,Z) \to (A, Z+1) + e^- +\bar{\nu}_e $  that proceed
through a conserved vector current to muon decay $\mu^- \to \nu_\mu e^- 
\bar{\nu}_e$. In the three-quark Cabibbo theory, this matrix element was
identified with $\cos \theta_C$.

(2) $|V_{us}|$: This is based on the analyses of the decays
 $K^+ \to \pi^0 \ell^+ 
\nu_\ell$ and $K^0 \to \pi^- \ell^+ \nu_\ell$ and beta decays of the hyperons. 
In the Cabibbo theory, this matrix element was identified with $\sin \theta_c$.

(3) $|V_{cd}|$: This is derived from the neutrino and antineutrino production
 of charm quarks from d quarks in a nucleon in  deep inelastic neutrino nucleon 
scattering experiments, $\nu_\ell + d \to \ell^- + c$. In the GIM-Cabibbo 
current, this matrix element is $\sin \theta_C$. 

(4) $|V_{cs}|$: This comes from the semileptonic decays of the charmed hadrons 
$D^\pm$ and $D^0$, involving for example the decay $D^\pm \to K^0 
\ell^\pm \nu_\ell$. Again, in the GIM-Cabibbo current, this matrix element
is identified with $\cos \theta_C$.

(5) $|V_{cb}|$: From the semileptonic decays of $B$ hadrons, such as
$\bar{B}^0 \to D^{*+} + \ell^{-} + \nu_\ell$, or the inclusive decay of
the $b$ quark $b \to c + \ell^{-} + \nu_\ell$.

(6) $|V_{ub}|$: Obtained from the semileptonic decays of $B$ hadrons into
 non-charmed
hadrons, such as $\bar{B}^0 \to \pi^{+} + \ell^{-} + \nu_\ell$, or
the inclusive semileptonic decays of a $b$ quark into a non-charm quark
$b \to u + \ell^{-} + \nu_\ell$.   

(7) $|V_{td}|$: From the measured mass difference between the mass eigenstates
 in the $B^0$ - $\overline{B^0}$ meson complex. Being an example of a FCNC 
process, this transition is a quantum effect and in the Standard Model
takes place through a box diagram very similar to the one shown in Fig.~1
for the $K^0$ - $\overline{K^0}$ system,
except that in this case the transition amplitude   
is dominated by the top quark due to its very large mass ($m_t \simeq 
175$ GeV). 

(8) $|V_{ts}|$: From the measured branching ratio of the electromagnetic
process $ b \to s + \gamma$, measured by the CLEO experiment at
CESR (Cornell) and recently also by the ALEPH collaboration at CERN.
Again, an example of a FCNC process, this
is also a quantum effect and again in the Standard Model the transition 
rate is dominated by the top quark. 

(9) $|V_{tb}|$: From the production and decay of the top quark in the
process $p\bar{p} \to t\bar{t} +X $ followed by the decay $t \to b + W^+$.

One sees that present knowledge of the matrix elements in the third
row of the CKM matrix involving the top quark in eq.~(\ref{eqvckmnum}), but 
also of the matrix 
elements $V_{cs}$, $V_{cd}$ and $V_{ub}$ is still rather imprecise.
A check of the unitarity of the CKM matrix  from the entries in
eq.~(\ref{eqvckmnum}) makes this quantitatively clear. Unitarity requires,
among other things, that the absolute squares of the elements in any row of
the CKM matrix add up to unity. We have at present
\begin{eqnarray}
|V_{ud}|^2 + |V_{us}|^2 + |V_{ub}|^2 &=& 0.997 \pm 0.002 ~, \nonumber\\
|V_{cd}|^2 + |V_{cs}|^2 + |V_{cb}|^2 &=& 1.18 \pm 0.33 ~, \nonumber\\
|V_{td}|^2 + |V_{ts}|^2 + |V_{tb}|^2 &=& 0.98 \pm 0.30 ~.
\end{eqnarray}
This shows that except for the first row, the information on the 
unitarity of the CKM matrix is very imprecise. However, all
data, within errors, are consistent with the CKM matrix being unitary.

\subsection{Unitarity Triangles}
\begin{figure}
\vskip -1.0truein
\centerline{\epsfxsize 3.5 truein \epsfbox {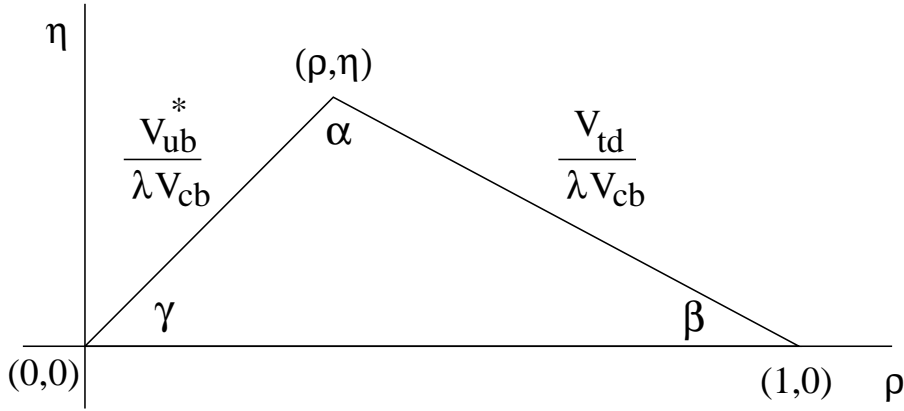}}
\vskip -1.2truein
%epsfig{file=rhoeta1.ps,bbllx=30pt,bblly=285pt,bburx=390pt,bbury=494pt,
%width=10cm}
\caption{The unitarity triangle. The angles $\alpha$, $\beta$ and $\gamma$
can be measured via CP violation in the $B$ system, and the sides
from the rates for various CC- and FCNC-induced $B$ decays.}
\label{triangle}
\end{figure}

The unitarity of the CKM matrix also requires that
any pair of rows or any pair of columns of this matrix be
orthogonal. This leads to six orthogonality conditions. These can be
depicted as triangles in the complex plane of the CKM parameter space.  
The constraint stemming from the orthogonality condition on the first and third
column of $V$,
\begin{equation}
    V_{ud} V_{ub}^* + V_{cd} V_{cb}^* + V_{td} V_{tb}^* = 0
\label{tdunit}
\end{equation}
is at the centre of contemporary theoretical and experimental attention. Since
present measurements are consistent with
$V_{ud} \simeq 1, ~V_{tb} \simeq 1$ and $V_{cd}\simeq -\lambda$, where
$\lambda = \sin \theta_C$, the unitarity relation (\ref{tdunit})
simplifies to: 
\begin{equation}
  V_{ub}^* + V_{td} \simeq -V_{cd} V_{cb}^* \simeq +\lambda V_{cb}^*,
\end{equation}
which can be conveniently depicted as a triangle relation in
the complex plane, as shown in Fig.~\ref{triangle}.
In drawing this triangle, we have used a 
representation of the CKM matrix due to Wolfenstein, characterized by four
 constants $A, \lambda, \rho$ and $\eta$. We have also  
rescaled the sides of the triangle by $\lambda 
V_{cb}$, which makes the base of the triangle real and of unit length and 
the apex of the triangle given by the point ($\rho,~\eta$) in the complex plane.
This is usually called the unitarity triangle (UT). Knowing the sides of the UT, 
the three angles of this triangle $\alpha, \beta$ and $\gamma$ are determined.
But these angles can also, in principle, be measured directly through
 observation of CP violation in various $B$ decays. 
By measuring both the sides and the angles, the UT
will be overconstrained, which is one of the principal goals of the current
and forthcoming experiments in flavour physics.

In the Wolfenstein representation, 
\begin{equation}
\label{cpasymdef}
\tan(\alpha) = \frac{\eta}{(\eta^2 - \rho(1-\rho))} ~,
~~~\tan (\beta) = \frac{\eta}{(1-\rho)} ~,
~~~\tan (\gamma) = \frac{\eta}{\rho}~.
\end{equation}

%
%This is Figure 2
\par
\begin{figure}
\vskip -1.0truein
\centerline{\epsfxsize 3.5 truein \epsfbox {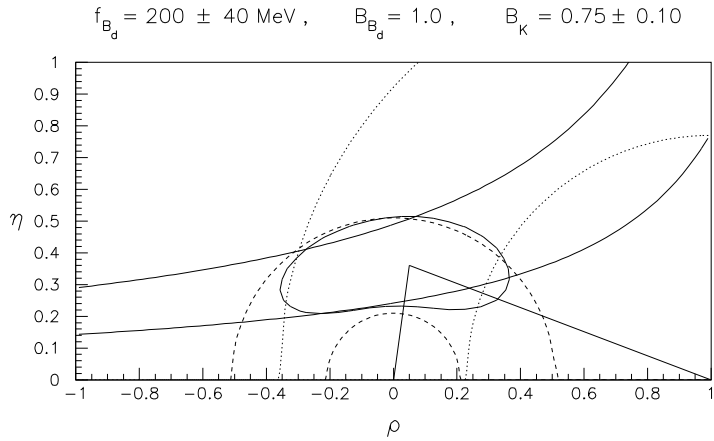}}
\vskip -1.2truein
%epsfig{file=rhoeta1.ps,bbllx=30pt,bblly=285pt,bburx=390pt,bbury=494pt,
%width=10cm}
\caption{ Allowed region in $\rho$-$\eta$ space obtained by overlaying
the individual constraints following from $|V_{ub}/V_{cb}|$ (dashed curves),
$\epsilon_K$ (solid curves), and $\Delta M_d$ (dotted curves), by
letting the hadronic quantities vary in the range shown above. The
$95\%$ C.L. contour resulting from a simultaneous fit of the data is
also shown (``Haggis"-type curve). The triangle shows the best fit.
}
\label{utfits}
\end{figure}

A profile of the UT based on our present knowledge of the CKM matrix is now 
given from which the CP-violating asymmetries which will be
measured in forthcoming experiments in B Physics can be estimated.
For this, the present experimental input can be summarized as
follows:
\begin{eqnarray}
\sqrt{\rho^2 + \eta^2} &=& 0.363 \pm 0.073~, \nonumber\\
(f_{B_d}\sqrt{\hat{B}_{B_d}}/1~\mbox{GeV}) \sqrt{(1-\rho)^2 + \eta^2} &=& 
0.202 \pm 0.017~,\nonumber\\
\hat{B}_K \eta [0.93 +(2.08 \pm 0.34)(1-\rho)] &=& 0.79\pm 0.11~,
\label{triangfit}
\end{eqnarray}
which come from the measurements of the CKM matrix element ratio 
$|V_{ub}/V_{cb}|=0.08 \pm 0.02$, the 
mass difference induced by the $B^0$ - $\overline{B^0}$ mixing, which is
measured very accurately, 
$\Delta M_d=(3.12 \pm 0.20)~\times~10^{-4}$ eV, and the CP-violating parameter
 in the $K^0$ - $\overline{K^0}$ system, $\epsilon_K =(2.28 \pm 0.013) \times 
10^{-3}$, which is likewise known very precisely.
 The quantities $f_{B_d}$, $\hat{B}_{B_d}$ and $\hat{B}_K$ are various
hadronic quantities whose knowledge is needed to analyze data. Present
estimates, based mostly on lattice QCD calculations, put them in the
range $f_{B_d}\sqrt{\hat{B}_{B_d}}= 200 \pm 40$ MeV and $\hat{B}_K=0.75 
\pm 0.10$. The resulting allowed regions in the
$(\rho,\eta)$ parameter space from each of these constraints individually and
the resulting overlap region from all the constraints put together 
are shown in Fig.~\ref{utfits}. The triangle drawn is
to guide the eye and represents the presently preferred solution. Two 
messages are clear: First, current theoretical uncertainties in hadronic 
quantities translate into rather large uncertainties in the profile of the 
unitarity triangle. Second, and despite this, a good part of the allowed 
parameter space
is now ruled out by data and  the CKM matrix provides a consistent
solution only over a limited parameter space.
%
%
%     This is Figure 3
\par
%
%
%  
%There are other impending constraints that one can put on the CKM 
%parameters. For
%example, the present lower bound on the $B_s^0$ - $\overline{B_s^0}$
%induced mixing, $\Delta M_s > 10.2~\mbox{(ps)}^{-1}$, arising from 
%experiments at LEP, removes some of the otherwise allowed negative-$\rho$
%region shown in Fig.~\ref{utfits}.

%
\subsection{CP Violation in $B$ Decays}

The paramount interest in $B$ physics lies in that it will test the 
CKM paradigm of CP violation in flavour-changing weak interactions.
The $B$ mesons can decay in many different ways, and a large number of
 their decay modes are potentially interesting from the point of view of
 observable CP-violating effects. In some of the decay modes, these effects
 can yield {\it clean} information, free of theoretical uncertainties, on the
 angles in the unitarity triangle of Fig.~\ref{triangle}. Since these angles
 are just the relative phases of various combinations of CKM elements, the
 clean information on the angles will stringently test the hypothesis
 that CKM phases cause CP violation.
The decay modes which can provide clean information on the angles include
 the decays of neutral $B$ mesons to final states which are CP eigenstates
 or at least can come from both a pure $B^0$ and a pure $\overline{B^0}$, and
 certain decays of charged $B$ 
 mesons. The decays of neutral $B$ mesons to CP eigenstates provide a
 particularly pretty example of how CP violation comes about, and how the
 phases of CKM matrix elements can be determined. In any decay, for CP
 violation to be non-zero, there must be interfering amplitudes with
  clashing phases. Now, in neutral $B$ decay to a CP eigenstate $f_{CP}$,
 there are two routes to the final state. If the parent $B$ was born as a
 $B^0$, it may (1) decay directly to $f_{CP}$, or else it may (2) turn via
 weak mixing
  into a $\overline{B^0}$, and then this $\overline{B^0}$ decays to
 $f_{CP}$. The amplitudes for these two routes must be added coherently,
 and will interfere.
 If the parent $B$ was born as a $\overline{B^0}$, decay to $f_{CP}$ can
 again proceed through two routes: $\overline{B^0}\to f_{CP}$, and
 $\overline{B^0} \to B^0 \to f_{CP}$. As before, the amplitudes for these
 two routes will interfere. If the CKM matrix
  elements have complex phases, then these (weak) amplitudes will have
 different phases than when the $B$ was born as a $B^0$. As a result, the
 interferences encountered in ($B$ born as $B^0$) $\to f_{CP}$ and ($B$
 born as $\overline{B^0}$) $\to f_{CP}$ will differ, and consequently
  the rates for these two decays will differ as well. Since the two decays
 are CP-mirror-image processes, the difference between their rates is a
 violation of CP.

 Since the rates $\Gamma$[$B$ born as $\optbar{B^0}$ $\to f_{CP}$ after
 time $t$] $\equiv \optbar{\Gamma}(t)$ depend nontrivially on the time $t$
 that the $B$ lives before decaying, experiments will study the
 time-dependent CP-violating asymmetry
\begin{equation}
    a_{f_{CP}}(t) \equiv \frac{\Gamma(t) - \overline{\Gamma}(t)}
    {\Gamma(t) + \overline{\Gamma}(t)}~.
\end{equation}
{\it When the unmixed $B$ decays, $B^0 \to f_{CP}$ and $\overline{B^0} \to
 f_{CP}$, are each dominated by one diagram}, $a_{f_{CP}}(t)$ is given by
 the simple expression
\begin{equation}
    a_{f_{CP}}(t) = \eta_{f_{CP}} \sin (\phi_{f_{CP}}) \sin (\Delta M t)~.
\label{eqC1}
\end{equation}

Here, $\eta_{f_{CP}} = \pm 1$ is the CP parity of the final state,
 $\Delta M$ is the mass difference between the two mass eigenstates of
 the $B^0 - \overline{B^0}$ system, and $\phi_{f_{CP}}$ is the phase of a
 certain product of CKM elements. Namely, $\phi_{f_{CP}}$ is the relative
 phase of the product of CKM elements to which the amplitude for
 $B^0 \to f_{CP}$ is proportional, and the product to which the amplitude for
 the alternate decay route, $B^0 \to \overline{B^0} \to
 f_{CP}$, is proportional. Of course, the identity of $\phi_{f_{CP}}$
 depends on the choice of $f_{CP}$.

 There are two neutral $B$ systems: $B^0_d(\bar{b}d)$ and its antiparticle,
 and $B^0_s(\bar{b}s)$ and its antiparticle. For the $B^0_d -
 \overline{B^0_d}$ system, the mass splitting $\Delta M_d$ between the mass
 eigenstates is already known, as previously mentioned. For the
  $B^0_s - \overline{B^0_s}$ system, the analogous splitting $\Delta M_s$
 will no doubt eventually be determined as well. Thus, the $\Delta M$ in
 Eq.~(\ref{eqC1}) for the CP asymmetry may be assumed known. For any chosen
 final state, the 
 CP parity $\eta_{f_{CP}}$ is also known. Thus, we see from
 Eq.~(\ref{eqC1}) that once the asymmetry $a_{f_{CP}}(t)$ is measured,
 $\sin (\phi_{f_{CP}})$ is {\it cleanly} determined, with no theoretical
 uncertainties. This makes it possible to cleanly test 
 whether complex phases of CKM matrix elements are indeed the origin of CP
 violation.

 As an example, suppose $f_{CP}$ is $J/\psi K_S$. In the decays
 ($B$ born as $\optbar{B^0_d}$) $\to J/\psi K_S$, each of the unmixed B
 decays, $B^0_d \to J/\psi K_S$ and $\overline{B^0_d} \to J/\psi K_S$,
{\it is} expected to be dominated by one diagram. Thus, Eq.~(\ref{eqC1})
 for $a_{f_{CP}}(t)$ should hold. The dominating diagrams are
 such that for this final state, $\phi_{f_{CP}}$ is simply $2\beta$, where
 $\beta$ is one of the angles in the unitarity triangle of
 Fig.~\ref{triangle}. Thus the decays ($B$ born as
  $\optbar{B^0_d}$) $\to J/\psi K_S$ can give us clean information on
 $\beta$. It appears that obtaining information on the other angles in the
 UT will be more difficult, but should still be possible. A major
 experimental effort will be made to determine all the angles of the UT.

How large are the CP-violating asymmetries in $B$ decays? They depend in part on
the mass-mixing related quantities $x_d \equiv \Delta M_d \cdot \tau(B_d)$ 
for the $B^0_d - \overline{B^0_d}$ system, which is well measured with 
$x_d \simeq 0.74$, and on 
 $x_s \equiv \Delta M_s \cdot \tau(B_s)$ for the $B^0_s - \overline{B^0_s}$
system, for which experiments at LEP (CERN) have provided only lower limits 
$x_s \geq 16$. Here, 
$\tau(B_d)(\tau(B_s))$ is the lifetime of the $B_d^0 (B_s^0)$ meson. But,
the CP asymmetries depend crucially on the angles of the UT, which can be 
estimated from the unitarity fits.
With the help of the relations given in eqs.~(\ref{cpasymdef}), the CP-violating 
asymmetries in $B$ decays can be expressed straightforwardly in terms
of the CKM parameters $\rho$ and $\eta$. The  constraints on
$\rho$ and $\eta$  discussed above can then be used to predict the
correlated ranges of the angles $\alpha$,
$\beta$ and $\gamma$ in the Standard Model. Representative of the current
theoretical expectations are the following 
ranges for the CP-violating rate asymmetries parametrized by $\sin
2\alpha$, $\sin 2\beta$ and $\sin^2 \gamma$, which are estimated by Ali \&
London in the context of the Standard Model at the end of 1997:
\begin{eqnarray}  
&~& -1.0 \leq \sin 2\alpha \le 1.0~, \nonumber \\
&~& 0.30 \leq \sin 2\beta \le 0.88~, \\
&~& 0.27 \leq \sin^2 \gamma \le 1.0~, \nonumber
\end{eqnarray}
with all ranges corresponding to 95\% C.L. (i.e., $\pm 2 \sigma$). The
currently preferred solutions of the unitarity fits yield: $\rho \simeq 0.12$
and $\eta \simeq 0.34$, which then translate into $\alpha \simeq 88^\circ$,
$\beta \simeq 21^\circ$ and $\gamma \simeq 72^\circ$. The central values
 of the parameters which determine the asymmetries are then: $\sin 2 \alpha
\simeq 0.07$, $\sin 2 \beta \simeq 0.67$ and $\sin^2 \gamma \simeq 0.89$. 
 These parameters will be measured in decays such as ($B$ born as
 $\optbar{B^0_d}$) $\to J/\psi K_S$, where the CP-violating asymmetry is
 proportional to $\sin 2 \beta$, ($B$ born as $\optbar{B^0_d}$) $\to \pi^+
 \pi^-$, which can determine $\sin 2 
 \alpha$, and ($B$ born as $\optbar{B^0_s}$) $\to D_s^\pm K^\mp$ or $B^\pm
 \to D K^\pm$, which can yield $\sin^2 \gamma$. The actual asymmetries in
the partial rates are expected to be quite large in some of these decays,
which will make them easier to measure in the next round of $B$ physics
experiments.  

 Additional decay modes which appear to be promising places to study CP
 violation include $B^\pm \to \pi^\pm K$, ($B$ born as $\optbar{B^0_d}$)
 $\to \pi^\pm K^\mp$, $B^\pm \to \pi^\pm \eta^\prime$,
($B$ born as $\optbar{B^0_d}$) $ \to K_S \eta^\prime$,
 ($B$ born as $\optbar{B^0_d}$) $\to D^{(*)\pm} \pi^\mp$, ($B$ born as
 $\optbar{B^0_d}$) $\to K^0 \overline{K^0}$, and many others. Moreover, one
expects measurable CP violation in the
inclusive radiative decays such as $B \to X_d + 
\gamma$, where $X_d$ is a system of light, non-strange, hadrons, and 
in exclusive radiative decays such as $B \to \rho + \gamma$, which are 
governed by the FCNC process $b \to d + \gamma$. These processes are
similar to the observed decays $B \to K^* + \gamma$ and $B \to X_s +
\gamma$, but are suppressed by about a factor 20. 
Measurements of CP asymmetries in these
processes do not directly determine the angles of the unitarity triangle.
However, they all depend on the parameters $\rho$ and $\eta$ and hence
their measurement will contribute to determine the UT more precisely,
and to the understanding of CP violation. However, most of these measurements 
will require sufficiently many $B$ hadrons that they will probably have to await 
the second round of experiments in $B$ factories at SLAC (Stanford),
KEK (Japan) and CESR (Cornell).

Apart from the CP violation measurements discussed above, some of  
the anticipated landmark measurements in $B$ physics include: (1) Determination 
of the mass splitting $\Delta M_s$ in the $B_s^0$ -$\overline{B_s^0}$ complex,
(2) Rare $B$ decays, such as $b \to d + \gamma$, $B \to
\rho^0(\omega) + \gamma$, $b \to s \ell^+ \ell^-$, $ b \to d \ell^+
\ell^-$ - all examples of FCNC processes, which have
been the driving force behind theoretical developments in flavour physics.

Likewise, several planned and ongoing experiments in $K$
physics will measure rare decays such as $K_L \to \pi^0 \nu \bar{\nu}$ and
$K^\pm \to \pi^\pm \nu \bar{\nu}$, and the CP-violating ratio  
 $\epsilon_K^\prime/\epsilon_K$. These important $K$-system measurements
 will complement the $B$-system experiments, and will help us to determine
 the properties of the unitarity triangle and to explore the origin of CP
 violation.
\section{Concluding remarks}

The elegant synthesis of seemingly diverse, and in their effective
strengths widely differing, empirical observations involving weak interactions 
 in terms of a universal constant $G_F$ and a $3 \times 3$ unitary matrix
is one of
the great simplifications in elementary particle physics. All data on weak
interactions can at present be analyzed and understood in terms of a few 
universal constants, and the consistency of the picture is indeed remarkable.
With improved theoretical and experimental precision, this consistency will 
provide in the future one of the most promising search strategies for finding
physics beyond the six quark Standard
Model of particle physics. A good candidate in that context is supersymmetry
which may contribute to many of the FCNC processes discussed here but
whose anticipated effects are quite subtle and their detection would require 
high precision data (see Chapter Ross).

Despite this success, there are many discomforting features which
 deserve attention. It must be stressed that the parameter $\epsilon_K$,
 which describes CP violation in $K^0 - \overline{K^0}$ mixing, and whose
 first measurement dates back some 35 years, still remains the only source
of information on CP
violation in laboratory experiments. This state of affairs is deeply disturbing,
in particular as CP violation  has a direct bearing on a fundamental phenomenon
in nature, namely the observed large-scale preponderance of matter over
antimatter in the universe. The next round of experiments in $B$ (and $K$)
physics will certainly help fill in some of the numerous blanks. At a deeper
level, however, the connection between complex phases in the CKM matrix and the
observed matter-antimatter asymmetry in the universe remains very much a matter
of speculation. It is conceivable that fundamental progress here may come from
completely different quarters, such as observation of CP violation in the lepton
sector and the understanding of baryo-genesis at the grand unification scale --
all aspects not directly related to the flavour physics of quarks discussed here.

We started this article with the discussion of the Fermi theory postulated
some sixty five years ago. The physics behind the effective
Fermi coupling constant, $G_F$, has come to be  understood in terms of a 
fundamental gauge interaction. The question is: Are the
elements of the CKM unitary matrix also some kind of effective parameters,
which some day one would be able to derive in terms of more fundamental
quantities? Some ideas along these lines are being pursued enthusiastically
in grand theoretical schemes where the CKM matrix elements are derived
in terms of quark masses. As theoretical and experimental
precision on the CKM matrix improves, many of these relations will
come under sharp experimental scrutiny. The emerging pattern
will help us to discard misleading theories, and perhaps 
single out a definitive and unique theoretical perspective.
The flavour problem - understanding the physics behind the parameters 
of the CKM matrix which seem to describe all flavour interactions at
present energies consistently - remains one of the most challenging 
problems of particle physics. 

\bigskip 
{\bf Bibliography:}

\begin{enumerate}

\item R.M.~Barnett et al. (Particle Data Group), Physical Review  D54 
(1996) 1 - 720.
\item {\it Leptons and Quarks}, L.B.~Okun, North Holland Physics Publishing,
Amsterdam, The Netherlands, 1987.
\item {\it CP Violation}, Advanced Series on Directions in High Energy 
Physics-Vol.~3, Editor: C.~Jarlskog, World Scientific Publishing Company,\\
Singapore, 1989.
\item {\it B Decays}, Revised 2nd Edition, Editor: Sheldon Stone, World
Scientific Publishing Company, Singapore, 1994.
%\item {\it CP Violation and Flavour Mixing in the
%Standard Model}, A.~Ali and D.~London, Nuclear Physics B(Proc.~Suppl.) 
%(1997) 297 -308.
%\item {\it Flavour Changing Neutral Current Processes and CKM
%Phenomenology}, A.~Ali, DESY Report DESY 97-256, hep-ph/9801270 (1998).
\end{enumerate}

\end{document}